\documentclass[a4paper, 11pt, oneside]{article}
\pdfoutput=1
\usepackage{float}
\usepackage{jheppub}
\usepackage{epstopdf}

\usepackage{bbm}   
\usepackage{amsmath}
\usepackage{graphicx}
\usepackage{cancel}
\usepackage{soul} 

\newcommand{\nc}{\newcommand}

\newlength{\absize}
\nc{\non}{\nonumber}
\nc{\hc}{\hbox {H.c.}} 
\nc{\noi}{\noindent}
\nc{\barx}{\bar{x}}
\nc{\pbarn}{\;\hbox {pb}}
\nc{\fbarn}{\;\hbox {fb}}
\newcommand{\bi}{\begin{itemize}}
\newcommand{\ei}{\end{itemize}}

\nc{\lsp}{\;\;\;\;\;}
\nc{\Lsp}{\;\;\;\;\;\;\;\;\;\;}  
\nc{\LLsp}{\lspace \lspace}
\nc{\lra}{\longrightarrow}
\nc{\beq}{\begin{equation}}  \nc{\eeq}{\end{equation}}
\nc{\bea}{\begin{eqnarray}}  \nc{\eea}{\end{eqnarray}}
\nc{\baa}{\begin{array}}     \nc{\eaa}{\end{array}}
\nc{\bit}{\begin{itemize}}   \nc{\eit}{\end{itemize}}
\nc{\ben}{\begin{enumerate}} \nc{\een}{\end{enumerate}}
\nc{\bce}{\begin{center}}    \nc{\ece}{\end{center}}
\nc{\bpm}{\begin{pmatrix}}   \nc{\epm}{\end{pmatrix}}
\nc{\bvt}{\begin{verbatim}}  \nc{\evt}{\end{verbatim}}
\def\lsim{\mathrel{\raise.3ex\hbox{$<$\kern-.75em\lower1ex\hbox{$\sim$}}}}
\def\gsim{\mathrel{\raise.3ex\hbox{$>$\kern-.75em\lower1ex\hbox{$\sim$}}}}

\def\lcal{{\cal L}}

\nc{\tanb}{\tan\beta}
\nc{\mch}{M_{H^\pm}}

\def\mch{M_{H^\pm}}

\nc{\for}{\lsp {\rm for} \lsp}
\nc{\andd}{\lsp {\rm and} \lsp}

\renewcommand{\Re}{\mbox{Re\thinspace}}
\renewcommand{\Im}{\mbox{Im\thinspace}}
\newcommand{\half}{{\textstyle\frac{1}{2}}}

\allowdisplaybreaks 

\def\i11{{\mathbbm 1}}

\begin{document}
\title{GOOFy - a systematic approach}
\author[a]{Bohdan Grzadkowski}
\emailAdd{bohdan.grzadkowski@fuw.edu.pl}
\affiliation[a]{Institute of Theoretical Physics, Faculty of Physics, University of Warsaw,\\ Pasteura 5, 02-093 Warsaw, Poland}
\author[b]{Odd Magne Ogreid}
\emailAdd{omo@hvl.no}
\affiliation[b]{Western Norway University of Applied Sciences, Postboks 7030, N-5020 Bergen, Norway}
\abstract{
{We investigate in detail a new class (GOOFy) of transformations for bosonic and fermionic fields that leave relevant parts of the Lagrangian density unchanged.}
The transformations act upon complex scalar fields $\Phi$ and $\Phi^\dagger$ employing generalized charge conjugation ($C$)  transformation in a {\it non-consistent manner}, i.e. allowing for $\Phi^\dagger \to (\Phi^\dagger)^\prime \neq (\Phi^\prime)^\dagger$. 
Requiring invariance of the kinetic terms under such transformations specifies the form of $(\Phi^\dagger)^\prime$.
An analogous strategy is also adopted for fermionic fields.
This offers a systematic way to construct new GOOFy-invariant field-theoretical models.
It turns out that theories which are invariant with respect to these GOOFy transformations satisfy relations among parameters that are found to be RGE-stable up to two and three loop orders, thus constituting fixed-points under running of the RGE. This has been verified for various theories containing different numbers of bosonic and fermionic fields. In particular it has been shown that the Standard Model~(SM) can not be a viable electroweak theory if demanding invariance under GOOFy transformations. However, the two-Higgs-Doublet Model (2HDM) may be invariant under GOOFy transformations (Yukawa couplings included), providing an interesting phenomenological example of physics beyond the SM. The most striking aspect of this study is the RGE stability of new relations between model parameters in a wide class of field theories. We also present a set of new relations between 2HDM potential parameters that constitute a fixed point under the running of the RGE up to at least three loop order.}
\maketitle

\section{Introduction}
This work is a continuation of the work initiated in \cite{Ferreira:2023dke}, in which we presented a class of {hitherto unknown transformations that pretend to act as symmetries} of the 2HDM, referred to as $r_0$ or GOOFy symmetries~\footnote{{The reader should note that we do not claim  the "GOOFy symmetries" are conventional, genuine symmetries in a sense of quantum field theory.}}. We demonstrated the existence of
(until then) unknown relations between parameters of the 2HDM scalar potential that form fixed points under the running of the renormalization group equations (RGE). Adopting the same notation, the most essential
aspect of the result found there was that the combined parameter relations
\beq
m_{11}^2+m_{22}^2=0, \lsp \lambda_1-\lambda_2=0 \lsp \text{and} \lsp \lambda_6+\lambda_7=0,
\label{fixpoint}
\eeq
form a fixed point under the running of RGE, i.e.~the corresponding beta functions also vanish, 
\bea
\beta (m_{11}^2+m_{22}^2) =0,\lsp  \beta(\lambda_1-\lambda_2) =0, \lsp \text{and} \lsp \beta(\lambda_6+\lambda_7)=0,
\eea when the conditions \eqref{fixpoint} are satisfied. 
More specifically, using the results of \cite{Bednyakov:2018cmx}, we were able to show that these relations are RGE stable up to all loop orders in perturbative expansion. The ideas developed in \cite{Ferreira:2023dke} have been further explored in \cite{Boh:2024MPI} and also by other authors in \cite{Haber:2025cbb,Trautner:2025yxz,Trautner:2025prm,deBoer:2025jhc}. The effective potential of GOOFy invariant models has been  discussed in \cite{Ferreira:2025ate,Pilaftsis:2024uub}.

In this work we shall formalize and extend the concept of GOOFy symmetries in a systematic and generalized manner, also to include fermionic degrees of freedom. It is important to recall the methodology developed in \cite{Ferreira:2023dke}. Decomposing the complex scalar fields into their real and imaginary parts, it was shown that there exists a transformation that implies the RGE-stable relations between model parameters, (\ref{fixpoint}). However, the transformation turns out to be a non-orthodox one, as it transforms real fields into purely imaginary ones and vice versa. In addition, invariance of the scalar kinetic terms 
requires an imaginary rescaling of coordinates $x^\mu \to i x^\mu$. In the present work, this coordinate transformation constitutes the single essential component in the very definition of generic GOOFy transformations. 

This work is structured as follows. In Section \ref{GOOFyscalars} we develop and define GOOFy transformations of scalars. Section \ref{GOOFyfermions} is devoted to extending GOOFy transformations to the  fermionic sector. Next, in Section \ref{Yukawa} 
we discuss constraints imposed by GOOFy transformations upon the Yukawa couplings in different models that contain various combinations of scalar and fermionic degrees of freedom. It is there shown that all constraints obtained from the GOOFy invariance are RGE stable up to two-loop order in perturbation theory.  Section \ref{SM} is devoted to the SM. We find that when GOOFy invariance is imposed upon the SM, the scalar mass term is not allowed, neither do we find any Yukawa matrices that are physically viable. On the other hand, in Section \ref{2HDM}, we show that the 2HDM provides a viable alternative to go beyond the SM of electroweak interactions. Our conclusions are presented in Section \ref{conc}.

\section{GOOFy transformation of scalars}
\label{GOOFyscalars}
The original GOOFy symmetry presented in \cite{Ferreira:2023dke} was a {transformation}  originating from the scalar sector of the 2HDM. We shall therefore start by generalizing the concept of GOOFy symmetries to a general complex scalar field theory.\footnote{{It is worth reminding for completeness, that in the Appendix A of \cite{Ferreira:2023dke} the GOOFy transformations of gauge fields for $SU(2) \times U(1)$ were discussed.  It has been shown there that in order to preserve the scalar kinetic terms in 2HDM, the gauge fields should transform according to $B_\mu \to i \, B_\mu$ and $W_{j\;\mu} \to (-1)^{j+1} \;i\; W_{j\;\mu}$, for $j=1,2,3$. Then it has been shown that the gauge kinetic terms are invariant.}} Let us for this purpose consider a scalar theory containing $N_\phi$ complex scalar fields organized into a  multiplet $\Phi^T\equiv (\phi_1, \cdots ,\phi_{N_\phi})$. 
In \cite{Ferreira:2023dke}, we showed that if the kinetic terms of the 2HDM Higgs bosons were to remain invariant under the $r_0$ or GOOFy transformation, one also had to include the imaginary scaling of coordinates, $x^\mu \to i x^\mu$ as part of the GOOFy transformation. 

In the present work we will take the complete opposite approach. We will start by {\it defining} a GOOFy transformation as a transformation of scalar, vector and fermionic fields ($\Phi, V_\mu,\psi$) that leave the corresponding kinetic terms invariant if we simultaneously transform the coordinates as $x^\mu \to i x^\mu$. When a GOOFy transformation leaves the kinetic terms invariant, we shall use the name GOOFy symmetry.

The kinetic terms of the model with $N_\phi$ complex fields can be written as $\lcal_\text{kin}=\left(\partial_\mu \Phi^\dagger\right) \left(\partial^\mu\Phi\right)$~\footnote{The most generic kinetic terms that might be considered are
$\lcal_\text{kin}=
a (\partial_\mu \Phi^\dagger) (\partial^\mu\Phi)+
\left[b (\partial_\mu \Phi^T) (\partial^\mu\Phi)+\hc\right]
$, where $a$ is real and $b$ is a complex number. However those terms will be disregarded in the present work.}. We shall  
allow for (potentially {\it non-consistent}\footnote{A transformation is called {\it non-consistent} if the transformed $\Phi^\dagger$ differs from Hermitian conjugation of transformed $\Phi$, i.e.
$(\Phi^\dagger)^\prime \neq (\Phi^\prime)^\dagger$. In other words, $\Phi^\dagger$ is transformed independently from $\Phi$. }) transformations of the following form\footnote{Note that the bar above $X_\phi$ does not mean complex conjugation in this context, but is used to denote a unitary matrix that is different from $X_\phi$. We use the star ($\star$) to denote complex conjugation in the present work.}, 
\bea
x^\mu \to i x^\mu, \lsp \Phi\to X_\phi \Phi^\star, \lsp \Phi^\dagger\to \Phi^T\bar X_\phi^\dagger,
\label{GOOFy_s}
\eea
where $X_\phi$ and $\bar X_\phi$ are both unitary $N_\phi \times N_\phi$ matrices. The transformation is inspired by a generic scalar $C$ transformation, see for instance \cite{Ferreira:2010bm}. Requiring $\lcal_\text{kin}$ to be invariant under the above transformation implies $\bar{X}_\phi=-X_\phi$. Then
\bea
x^\mu \to i x^\mu, \lsp \Phi\to X_\phi \Phi^\star, \lsp \Phi^\dagger\to -\Phi^T X_\phi^\dagger
\label{eq:trafo_rules}
\eea
will constitute the GOOFy transformation of our model, leaving $\lcal_\text{kin}$ invariant.
Note that, since product of derivatives of fields changes sign, a {\it consistent} transformation, i.e. such that $(\phi^\dagger)^\prime = (\phi^\prime)^\dagger$, can not be a symmetry of the kinetic terms. In order to make the kinetic terms invariant, an independent transformation (provided by $\bar{X}_\phi=-X_\phi$) is necessary.   
It is also worth noticing that the transformation matrix $X_\phi$ can not be reduced to the identity matrix by adopting the freedom of unitary field redefinition. In order to do so one would need to replace $\Phi$ by $U\Phi^\prime$ with $U$ being a unitary matrix such that $X_\phi=UU^T$. Then it is seen that only symmetric $X_\phi$ could be transformed into the identity matrix.

Next, let us consider the scalar potential starting with the following mass terms
\bea
V(\Phi)=\Phi^\dagger M^2\Phi+(\Phi^T\mu^2\Phi +\hc),
\eea
where Hermiticity of the Lagrangian implies that $M^2$ is a Hermitian $N_\phi \times N_\phi$ matrix, while by construction, $\mu^2$ is a symmetric complex $N_\phi \times N_\phi$ matrix. 
Demanding invariance of the $\Phi^\dagger \Phi$ mass terms in a general basis requires
\bea
	M^2=-X_\phi^T \; \left(M^2\right)^T\; X_\phi^\star.
\label{Msquaregeneral}
\eea
It immediately follows that $\text{Tr}\, M^2 = 0$. This requirement forbids this kind of mass term in the case of a single complex scalar field ($N_\phi=1$). For $N_\phi\geq 2$, this type of mass terms is allowed.

Unless specified otherwise, we will hereafter choose a basis such that $M^2$ is real and diagonal. Alternatively, since $\mu^2$ is symmetric, it can be diagonalized\footnote{We cannot simultaneously diagonalize $M^2$ and $\mu^2$. Therefore, when we consider the simultaneous presence of both types of mass terms, we will assume only $M^2$ to be diagonal.} by a change of basis into a non-negative diagonal matrix\footnote{This results follows from Corollary 2.6.6a of \cite{Horn_Johnson_2012}.}.
The invariance of the $\Phi^\dagger \Phi$ mass term in a basis with symmetric and real $M^2$ matrix reads
\bea
M^2=-X_\phi^T \; M^2\; X_\phi^\star,
\label{Msquare}
\eea
which could alternatively be formulated as $\{M^2,X_\phi\}=0$. 
Since $\det X_\phi\neq 0$, it follows that if $\det M^2\neq 0$, then $N_\phi=2k$ (even). If instead $\det M^2 = 0$, then
$N_\phi$ is unconstrained.

{There is a comment here in order. If we consider unitary (consistent) transformations (Higgs Family transformations) on the scalar multiplet $\Phi$, it will be of the form $\Phi\to X_\phi\Phi$ and $\Phi^\dagger\to \Phi^\dagger X_\phi^\dagger$. Invariance of the scalar mass terms will then imply
\bea
M^2 = X_\phi^\dagger M^2 X_\phi.
\label{uni_inv}
\eea 
}
{
If we consider anti-unitary (consistent) transformations (generalized CP transformations) on the scalar multiplet $\Phi$, it will be of the form $\Phi\to X_\phi\Phi^\star$ and $\Phi^\dagger\to \Phi^T X_\phi^\dagger$. Invariance of the scalar mass terms will imply
\bea
M^2 = X_\phi^T \left(M^2\right)^T X_\phi^*.
\label{anti-uni_inv}
\eea 
Then taking the trace on both sides of either \eqref{uni_inv} or \eqref{anti-uni_inv}  we see that in both cases, we get no constraint on $\text{Tr}\, M^2$. That demonstrates that the basis invariant constraint $\text{Tr}\, M^2 = 0$ that follows from demanding invariance of the scalar mass terms under GOOFy transformations is indeed unique to GOOFy transformations, and cannot be a result of invariance under the familiar unitary or anti-unitary transformations. } 

Now we are ready to demand invariance of the $\Phi^T\Phi$ mass term\footnote{ We have explicit transformation rules for $\Phi$ and $\Phi^\dagger$ in (\ref{eq:trafo_rules}), but there are ambiguities in how $\Phi^\star$ and $\Phi^T$ should transform, since 
\beq
\Phi^\star=
\left\{ 
   \begin{array}{c}
   (\Phi^\dagger)^T \to - X_\phi^\star\Phi\\
   (\Phi)^\star \to  + X_\phi^\star\Phi
  \end{array}
\right.
\lsp \text{and} \lsp
\Phi^T=
\left\{ 
   \begin{array}{c}
   (\Phi^\dagger)^\star \to - \Phi^\dagger X_\phi^T\\
   (\Phi)^T \to + \Phi^\dagger X_\phi^T
  \end{array}.
\right.\nonumber 
\eeq
Upon deriving eq. (\ref{musquare}), we have used 
$\Phi^T=(\Phi)^T\to \Phi^\dagger X_\phi^T$ and $\Phi^\star=(\Phi^\dagger)^T\to -X_\phi^\star \Phi$. Had we instead adopted $\Phi^T=(\Phi^\dagger)^\star\to -\Phi^\dagger X_\phi^T$ and $\Phi^\star=(\Phi)^\star\to X_\phi^\star \Phi$, we would have ended up with the condition $\left(\mu^2\right)^\star = - X_\phi^T\;\mu^2\; X_\phi$ for invariance. For remaining two transformation options $\mu^2$ must vanish.\label{footnote:musq}},
\bea
\left(\mu^2\right)^\star = X_\phi^T\;\mu^2\; X_\phi.
\label{musquare}
\eea
It is easy to see that this kind of a mass term is allowed even for a one-dimensional complex field.

Let us now illustrate the implications of demanding invariance for one or both types of mass terms in models with two complex scalar fields.
\subsection{$\Phi^\dagger\Phi$ mass only}
We assume here that $N_\phi=2$. 
For $X_\phi$ we consider the most general $U(2)$ matrix 
\bea
X_\phi=e^{i\psi}\left(
\begin{array}{cc}\cos\theta & e^{-i\xi}\sin\theta\\ -e^{i\chi}\sin\theta & e^{i(\chi-\xi)}\cos\theta
\end{array}\right).
\eea
In a general basis, using the invariance condition \eqref{Msquaregeneral}, we find the following form of $X_\phi$,
\bea
X_\phi= 
\left( {\begin{array}{cc}
		0 & e^{i\theta_1} \\
		-e^{i\theta_1} & 0 \\
\end{array} } \right)
\label{exphigeneral}
\eea
and the $\Phi^\dagger \Phi$ mass matrix is of the form
\bea
M^2= 
\left( {\begin{array}{cc}
		m_{11}^2 & m_{12}^2  \\
		(m_{12}^2)^\star & -m_{11}^2 \\
\end{array} } \right).
\label{massqgeneral}
\eea
In the real and diagonal basis where $m_{12}^2=0$, there is an extra degree of freedom in $X_\phi$, which now reads\footnote{Note that the matrix $X_\phi$ of (\ref{exphigeneral}) can be considered a special case of (\ref{exphi}) with $\theta_2=\theta_1+\pi$} 
\bea
X_\phi= 
  \left( {\begin{array}{cc}
   0 & e^{i\theta_1} \\
   e^{i\theta_2} & 0 \\
  \end{array} } \right)
\label{exphi}
\eea
and the $\Phi^\dagger \Phi$ mass matrix is of the form\footnote{Note that the conclusion $ \mathrm{Tr} M^2=0$ is basis independent.}
\bea
M^2= 
  \left( {\begin{array}{cc}
   m_{11}^2 & 0  \\
   0 & -m_{11}^2 \\
  \end{array} } \right).
\label{massq}
\eea
It is worth noting that $M^2$ does not depend on $\theta_{1,2}$, it is independent of $X_\phi$.
Note also the similarity of the $\Phi^\dagger\Phi$ mass matrices allowed here and the GOOFy constraint upon scalar masses ($m_{11}^2+m_{22}^2=0$)
found for 2HDM  in \cite{Ferreira:2023dke}. Since scalars appear in 2HDM in terms of two $SU(2)$ doublets therefore the 2HDM, even though it contains 8 real scalar fields, does match the constraints of the $N_\phi=2$ model with $\Phi^\dagger M^2\Phi$ mass terms discussed here.
\subsection{$\Phi^T\Phi$ mass only}
The condition \ref{musquare} is rather easy to satisfy, e.g. it would hold for any real and symmetric $\mu^2$  matrix (the case considered here) if $X_\phi=\mathbb{I}_{N_\phi}$ for any $N_\phi$.
\subsection{Coexistence of $\Phi^\dagger\Phi$ and $\Phi^T\Phi$ mass terms}

In this case we have to satisfy both \eqref{Msquare} and \eqref{musquare}. Hereafter we adopt the basis in which $M^2$ is real and diagonal for $N_\phi=2$. Therefore
$X_\phi$ has a generic form of \eqref{exphi} with additional constraints emerging from  \eqref{musquare}.
The $\mu^2$ mass term and $X_\phi$ read 
\bea
\mu^2=
\left( {\begin{array}{cc}
		\mu^2_{11} & (-1)^n i^m |\mu_{12}^2|e^{-i\theta_1} \\
		(-1)^n i^m |\mu_{12}^2|e^{-i\theta_1} & (\mu^2_{11})^\star e^{-2i\theta_1} \\
\end{array} } \right)
\lsp \text{with} \lsp
X_\phi=e^{i\theta_{1}}\left( {\begin{array}{cc}
		0 & 1 \\
		(-1)^m & 0 \\
\end{array} } \right)\nonumber\\
\eea
and
\bea
\mu^2=(-1)^n |\mu_{12}^2|e^{-i(\theta_1+\theta_2)/2}
\left( {\begin{array}{cc}
		0 & 1 \\
		1 & 0 \\
\end{array} } \right)
\lsp \text{with} \lsp
X_\phi= 
\left( {\begin{array}{cc}
		0 & e^{i\theta_1} \\
		e^{i\theta_2} & 0 \\
\end{array} } \right)\nonumber , \\
\eea
where $m,n$ are integer numbers.

\section{GOOFy transformation of fermions}
\label{GOOFyfermions}
For fermions we also start with $C$-inspired transformation:
\bea
x^\mu \to i x^\mu, \lsp \Psi\to -X_\psi \gamma_0 C \Psi^\star, \lsp \bar\Psi \to -\Psi^T C^{-1} \bar X_\psi^\dagger
\label{eq:goofyfermions}
\eea
where $\Psi^T\equiv (\psi_1,\cdots, \psi_{N_\psi})$ denotes a multiplet of $N_\psi$  fermionic fields while $X_\psi$ and $\bar X_\psi$ are unitary $N_\psi \times N_\psi$ matrices. Again, we allow for independent field transformations of $\psi$ and $\bar\psi$. GOOFy symmetries are still defined by demanding invariance of $\lcal_\text{kin}=i \bar\Psi \gamma^\mu \partial_\mu \Psi$ under (\ref{eq:goofyfermions}), which in turn implies $\bar X_\psi=-iX_\psi$ i.e.
\bea
x^\mu \to i x^\mu, \lsp \Psi\to -X_\psi \gamma_0 C \Psi^\star, \lsp \bar\Psi \to -\Psi^T C^{-1} i X_\psi^\dagger,
\label{fer_trans}
\eea
which will therefore constitute the GOOFy transformation for fermionic fields.
\subsection{Dirac mass}
In order to preserve the Dirac fermionic mass term ${\cal L}_\text{D-mass}=-\bar\Psi M_D \Psi$ (with $M_D^\dagger=M_D$) under the GOOFy transformation \eqref{fer_trans},
the following condition must hold
\bea
M_D=i X_\psi^T M_D^\star X_\psi^\star.
\label{Dir_mass}
\eea
Since $M_D$ is Hermitian, it turns out, that the above equation implies $M_D=0$ regardless of $N_\psi$ for any unitary $X_\psi$. We conclude that fermionic Dirac mass terms are not compatible with GOOFy symmetry. 

\subsection{Majorana mass}
In order to preserve the Majorana mass term ${\cal L}_\text{M-mass}=-1/2\left( \overline{\Psi^C} M_M \Psi +\bar\Psi M_M^\dagger \Psi^C \right)$ under the GOOFy transformation \eqref{fer_trans}, the following condition must hold\footnote{ We have explicit transformation rules for $\Psi$ and $\bar\Psi$ in \eqref{fer_trans}, but there are ambiguities in how $\Psi^C$ and $\overline{\Psi^C}$ should transform, since 
\beq
\psi^C=
\left\{ 
   \begin{array}{c}
   C(\bar\Psi)^T \to i X_\psi^\star\Psi \\
   C\gamma_0(\Psi)^\star \to X_\psi^\star\Psi
  \end{array}
\right.
\lsp \text{and} \lsp
\overline{\psi^C}=
\left\{ 
   \begin{array}{c}
   -(\Psi)^TC^{-1} \to \bar\Psi X_\psi^T\\
   (\bar\Psi)^\star C^{-1}\gamma_0\to -i\bar\Psi X_\psi^T
  \end{array}
\right. 
\non 
\eeq
Upon deriving eq. \eqref{Maj_mass}, we have used 
$\Psi^C=C\gamma_0(\Psi)^\star \to X_\psi^\star\Psi$ and $\overline{\Psi^C}= (\bar\Psi)^\star C^{-1}\gamma_0\to -i\bar\Psi X_\psi^T$. Had we instead adopted another transformation option, $M_M$ would have to vanish.
}
\bea
M_M=i X_\psi^\dagger M_M^\dagger X_\psi^\star.
\label{Maj_mass}
\eea
In addition, anticommutation of fermionic fields implies that $M_M$ is a symmetric matrix, i.e. $M_M^T=M_M$. Then
\bea
M_M=i X_\psi^\dagger M_M^\star X_\psi^\star.
\label{Maj_mass2}
\eea
Note the similarity between Dirac $\bar\Psi M_D \Psi$ and Majorana  $(\overline{\Psi^C} M_M \Psi + \hc )$ mass terms and scalar mass terms $\Phi^\dagger M^2 \Phi$ and $(\Phi^T \mu^2 \Phi +\hc )$, respectively.

\section{Yukawa interactions}
\label{Yukawa}
In this section we are going to investigate constraints imposed upon Yukawa couplings by demanding invariance with respect to the GOOFy transformations we have established.
 
We are going to consider the following general~\footnote{There is also an analogous pseudo-scalar coupling $-(\bar \Psi \Gamma \Phi \gamma_5 \Psi +\hc)$ and a Majorana-like coupling $ -1/2( \overline{\Psi^C} \Phi \Psi + \hc)$ which will not be discussed here. } Yukawa interactions:
\beq
-\lcal_Y=\bar\Psi \Gamma \Phi \Psi+\bar\Psi \Phi^\dagger \Gamma^\dagger \Psi \equiv
\bar\psi_a (\Gamma_i)_{ab} \phi_i \psi_b+\bar\psi_a (\phi_i)^\star (\Gamma_i)_{ba}^\star \psi_b,
\label{Yukawa_coupl}
\eeq
where the letters from the beginning of the alphabet $a,b=1,2, \cdots$ denote fermionic flavours while those starting in the middle denote complex scalars, so $i=1,2, \cdots$. 

Requiring invariance of $\lcal_Y$ under GOOFy transformations \eqref{fer_trans} and \eqref{eq:trafo_rules} implies the following condition:

\beq
(\Gamma_i)^\star=i \; X_\psi^T  \Gamma_j X_\psi^\star \; (X_\phi)_{ji}
\label{con_gen}
\eeq
These equations constitute a fixed point under the running of the RGE if and only if they are order by order scale invariant, i.e.
\bea
\beta^{(n)}\left[(\Gamma_i)^\star\right]=i \; X_\psi^T  \beta^{(n)}\left(\Gamma_j\right) X_\psi^\star \; (X_\phi)_{ji}.
\label{scal_inv}
\eea
Here, $\beta^{(n)}(\Gamma_j)$ denotes the $n$-loop beta function for the Yukawa matrix $\Gamma_j$. The
relation \eqref{con_gen} potentially correspond to hitherto unknown symmetries. {Below we present several models and we check RGE invariance of the relation \eqref{con_gen} together with \eqref{Msquaregeneral} or \eqref{musquare}. However, we will limit ourself to scalar potentials that contain only mass terms.
Therefore variables (model parameters) contributing to beta functions in this case are mass parameters and Yukawa couplings. Contributions containing quartic couplings, if included, would cancel in running equations independently. }

\subsection{$X_\psi=\mathbb{I}_{N_\psi}$ and $X_\phi=\mathbb{I}_{N_\phi}$}
For the simplest, though non-trivial, GOOFy transformation \eqref{con_gen} with $X_\psi=\mathbb{I}_{N_\psi}$ and $X_\phi=\mathbb{I}_{N_\phi}$ one obtains
\beq
\Gamma^\star = i \Gamma,
\label{con_one}
\eeq
which is equivalent to $\Re{\Gamma}+\Im{\Gamma}=0$. Note that $\Gamma$ is a $N_\psi\times N_\psi$ complex matrix carrying also a Higgs index, i.e. $(\Gamma_i)_{ab}$ with $i=1,\cdots, N_\phi$. Using results obtained with PyR@TE3 \cite{Sartore:2020gou} to calculate beta functions of the Yukawa couplings, we have explicitly verified that
\beq
\beta^{(1)}(\Gamma^\star) = i \beta^{(1)}(\Gamma)
\eeq 
for $N_\phi=1,\ldots,10$ and that 
\beq
\beta^{(2)}(\Gamma^\star) = i \beta^{(2)}(\Gamma)
\eeq
for $N_\phi=1,\ldots,5$, demonstrating RGE-stability of \eqref{con_one} at both one-loop and two-loop order for certain finite $N_\phi$ in agreement with expectations for a GOOFy-symmetric theory. We conjecture that this will be the case for any finite value of $N_\phi$.

In this case \eqref{Msquare} implies $M^2=0$, so this is not a viable option for an electroweak theory. Therefore
we employ the freedom of choosing a basis in which the matrix $\mu^2$ is real (and non-negative). 
Then eq.~\eqref{musquare} implies
\bea
(\mu^2)^\star=\mu^2,
\label{eq:musq}
\eea
equivalent to $\mu^2$ being real, which is already satisfied in our chosen basis\footnote{Had we instead chosen the option described in Footnote \ref{footnote:musq} leading to $(\mu^2)^\star=-\mu^2$, we would conclude that $\mu^2$ is purely imaginary, which is in conflict with our assumption of a real $\mu^2$.}. 
Using results obtained with PyR@TE3 we find that eq.~\eqref{eq:musq} is RGE-stable up to two-loop order for $N_\phi=1,\ldots 4$, provided also eq.~\eqref{con_one} is satisfied.

\subsection{$X_\phi=\mathbb{I}_{N_\phi}$, $N_\psi=3$}
For $X_\phi=\mathbb{I}_{N_\phi}$
\bea
\Gamma_j^\star=i X_\psi^T \Gamma_j  X_\psi^\star
\label{con_two}
\eea
In this case again \eqref{Msquare} implies $M^2=0$, however, according to \eqref{musquare}, $\mu^2$ might be a non-zero Hermitian matrix, which we can choose to be diagonal and non-negative.  Then one can solve for $\Gamma$ using the standard form for generalized $CP$ transformations found in \cite{Ecker:1987qp}, e.g. for $N_\psi=3$
\bea
X_\psi= 
  \left( {\begin{array}{ccc}
   \cos\theta_\psi & \sin\theta_\psi & 0\\
  -\sin\theta_\psi & \cos\theta_\psi & 0\\
        0          &       0         & 1\\
  \end{array} } \right)
  \lsp \text{and} \lsp
\Gamma_j=(i-1)
  \left( {\begin{array}{ccc}
   a_{11} & a_{12} &    0\\
   -a_{12} &   a_{11} &    0\\
     0   &    0    & a_{33}\
  \end{array} } \right),
\label{XA1}
\eea
where $a_{ij}$ are real parameters while $j=1, \dots , N_\phi$.
For $\theta_\psi=\pi/2$ there is also the solution
\bea
X_\psi= 
\left( {\begin{array}{ccc}
		0 & 1 & 0\\
		-1 & 0 & 0\\
		0          &       0         & 1\\
\end{array} } \right)
\lsp \text{and} \lsp
\Gamma_j=
\left( {\begin{array}{ccc}
		z_{11} & z_{12} &    0\\
		iz_{12}^\star &  -iz_{11}^\star  &    0\\
		0   &    0    & (i-1)r\\
\end{array} } \right),
\label{XA2}
\eea
where $z_{ij}$ are complex, whereas $r$ is real.

Using the beta functions obtained by PyR@TE3, we verify that \eqref{con_two} is RGE-stable up to two-loop order for $X_\psi$
 and $\Gamma_j$ of the form given both as in \eqref{XA1} and as in \eqref{XA2} for $N_\phi=1,\ldots 4$.

Also in this case \eqref{Msquare} implies $M^2=0$, and again we choose a basis in which the matrix $\mu^2$ is real (and non-negative), leading to \eqref{eq:musq}. 
Using results obtained with PyR@TE3 we find that eq.~\eqref{eq:musq} is RGE-stable up to two-loop order for $X_\psi$
and $\Gamma_j$ of the form given both as in \eqref{XA1} and as in \eqref{XA2} for $N_\phi=1,\ldots 4$.
\subsection{$X_\psi=\mathbb{I}_{N_\psi}$, $N_\phi=2$}
For 
$X_\psi=\mathbb{I}_{N_\psi}$
\bea
\Gamma_i^\star=i \Gamma_j (X_\phi)_{ji}
\label{con_three}
\eea
In this case \eqref{Msquare} allows for $M^2 \neq 0$ if $X_\phi \neq \mathbb{I}_{N_\phi}$. In addition we assume $\mu^2=0$, therefore \eqref{exphi} holds. Then, for a fixed $X_\phi$, one can solve \eqref{con_three} for $\Gamma$.
For instance for $N_\phi=2$ one finds that if $M_{11}^2, M_{22}^2 \neq 0$ only $\theta_2=\theta_1+2\pi n$ is allowed with $\Gamma_2=-ie^{-i\theta_1} \Gamma_1^\star$ i.e.
\bea
X_\phi= e^{i \theta_1} 
  \left( {\begin{array}{cc}
   0 & 1 \\
   1 & 0 \\
  \end{array} } \right)
   \lsp \text{with} \lsp 
\Gamma = 
  \left( {\begin{array}{cc}
   \Gamma_1,  &  -i e^{-i\theta_1} \Gamma_1^\star  
  \end{array} } \right).
\label{exphi2}
\eea
Using the beta functions obtained by PyR@TE3, we verify that \eqref{con_three} is RGE-stable up to two-loop order for $X_\phi$ and $\Gamma_j$ of the form given in \eqref{exphi2}.

Furthermore, we find that \eqref{Msquare} is RGE-stable up to two-loop order for $M^2$ of the form given in \eqref{massq} when $X_\phi$ and $\Gamma_j$ are of the form given in \eqref{exphi2}.

\subsection{$N_\phi=2$, $N_\psi=2$}
We have also investigated the case $N_\phi=2$ and $N_\psi=2$.
For $X_\phi$ of the form of \eqref{exphi}, eq.~\eqref{con_gen} yields
\bea
\Gamma_1^\star=i e^{i\theta_2}\; X_\psi^T  \Gamma_2 X_\psi^\star,\label{eq:A1star}\\
\Gamma_2^\star=i e^{i\theta_1}\; X_\psi^T  \Gamma_1 X_\psi^\star.\label{eq:A2star}
\eea	
The above equations could be solved for $\Gamma_{1,2}$, e.g. first one finds an equation for $\Gamma_1$ 
\bea
\Gamma_1=e^{-i(\theta_2-\theta_1)}\; X_\psi^\dagger   X_\psi^T \Gamma_1 X_\psi^\star X_\psi,
\label{eq:A1}
\eea	
which then allows to find a solution for $\Gamma_2$.
We choose $X_\psi$ to be a real orthogonal $2\times 2$ matrix parameterized by an angle $\theta$, i.e.
\bea
X_\psi = 
\left( {\begin{array}{cc}
		\cos\theta & \sin\theta \\
		-\sin\theta & \cos\theta 
\end{array} } \right).
\eea
Non-trivial solutions exist whenever the determinant of the coefficient matrix of the resulting linear system of equations derived from \eqref{eq:A1} for entries of $\Gamma_1$ vanish, i.e.
\bea
64\left[\cos(\theta_1-\theta_2)-\cos(4\theta)\right]^2\left[\sin\left( \frac{\theta_1-\theta_2}{2}\right)\right]^4=0,
\eea
with solutions $\theta_2=\theta_1$ or $\theta_2=\theta_1\pm 4\theta$. We find six different non-trivial solutions of \eqref{eq:A1}.
\\
{\it Solution 1:}
\bea
X_\phi = e^{i \theta_1}
\left( {\begin{array}{cc}
		0 & 1 \\
		1 & 0 
\end{array} } \right),\;
X_\psi = 
\left( {\begin{array}{cc}
		1 & 0 \\
		0 & 1 
\end{array} } \right),\;
\Gamma_1=\begin{pmatrix}
	z_{11} & z_{12} \\
	z_{21} & z_{22}
	\end{pmatrix},\;
\Gamma_2= -ie^{-i\theta_1} 
\begin{pmatrix}
	z_{11}^\star & z_{12}^\star \\
	z_{21}^\star & z_{22}^\star
\end{pmatrix}.\nonumber\\
\eea
{\it Solution 2:}
\bea
&X_\phi = e^{i \theta_1}
\left( {\begin{array}{cc}
		0 & 1 \\
		1 & 0 
\end{array} } \right),\;
X_\psi = 
\left( {\begin{array}{cc}
		0 & 1 \\
		-1 & 0
\end{array} } \right),\nonumber\\
&\Gamma_1=\begin{pmatrix}
	z_{11} & z_{12} \\
	z_{21} & z_{22}
\end{pmatrix},\;
\Gamma_2= -ie^{-i\theta_1} 
\begin{pmatrix}
	z_{22}^\star & -z_{21}^\star \\
	-z_{12}^\star & z_{11}^\star
\end{pmatrix}.
\eea
{\it Solution 3:}
\bea
&X_\phi = e^{i \theta_1}
\left( {\begin{array}{cc}
		0 & 1 \\
		1 & 0 
\end{array} } \right),\;
X_\psi = 
\left( {\begin{array}{cc}
\cos\theta & \sin\theta \\
-\sin\theta & \cos\theta 
\end{array} } \right),\nonumber\\
&\Gamma_1=\begin{pmatrix}
	z_{11} & z_{12} \\
	-z_{12} & z_{11}
\end{pmatrix},\;
\Gamma_2= -ie^{-i\theta_1} 
\begin{pmatrix}
	z_{11}^\star & z_{12}^\star \\
	-z_{12}^\star & z_{11}^\star
\end{pmatrix}.
\eea
\
{\it Solution 4:}
\bea
&X_\phi = e^{i \theta_1}
\left( {\begin{array}{cc}
		0 & 1 \\
		-1 & 0 
\end{array} } \right),\;
X_\psi = 
\left( {\begin{array}{cc}
		\frac{1}{\sqrt{2}} & \frac{1}{\sqrt{2}} \\
		-\frac{1}{\sqrt{2}} & \frac{1}{\sqrt{2}} 
\end{array} } \right),\nonumber\\
&\Gamma_1=\begin{pmatrix}
	z_{11} & z_{12} \\
	z_{12} & -z_{11}
\end{pmatrix},\;
\Gamma_2= -ie^{-i\theta_1} 
\begin{pmatrix}
	-z_{12}^\star & z_{11}^\star \\
	z_{11}^\star & z_{12}^\star
\end{pmatrix}.
\eea
{\it Solution 5:}
\bea
&X_\phi = e^{i \theta_1}
\left( {\begin{array}{cc}
		0 & 1 \\
		e^{4i\theta} & 0 
\end{array} } \right),\;
X_\psi = 
\left( {\begin{array}{cc}
\cos\theta & \sin\theta \\
-\sin\theta & \cos\theta 
\end{array} } \right),\nonumber\\
&
\Gamma_1=z_{11}\begin{pmatrix}
	1 & -i \\
	-i & -1
\end{pmatrix},\;
\Gamma_2= -ie^{-i(\theta_1+2\theta)}z_{11}^\star
\begin{pmatrix}
	1 & i \\
	i & -1
\end{pmatrix}.
\eea
{\it Solution 6:}
\bea
&X_\phi = e^{i \theta_1}
\left( {\begin{array}{cc}
		0 & 1 \\
		e^{-4i\theta} & 0 
\end{array} } \right),\;
X_\psi = 
\left( {\begin{array}{cc}
		\cos\theta & \sin\theta \\
		-\sin\theta & \cos\theta 
\end{array} } \right),\nonumber\\
&
\Gamma_1=z_{11}\begin{pmatrix}
	1 & i \\
	i & -1
\end{pmatrix},\;
\Gamma_2= -ie^{-i(\theta_1-2\theta)}z_{11}^\star
\begin{pmatrix}
	1 & -i \\
	-i & -1
\end{pmatrix}.
\eea
For all solutions, $z_{ij}$ are arbitrary complex parameters.

Using the beta functions obtained by PyR@TE3, we verify that \eqref{eq:A1star} and \eqref{eq:A2star} are RGE-stable up to two-loop order for $X_\phi$, $X_\psi$ and $\Gamma_j$ of the form given in all six solutions.

Furthermore, we find that eq.~\eqref{Msquare} is RGE-stable up to two-loop order for $M^2$ of the form given in \eqref{massq} for $X_\phi$, $X_\psi$ and $\Gamma_j$ of the form given in all six solutions.

In all cases considered above, GOOFy constraints on Yukawa matrices turns out to be
RGE stable for various choices for the number of fermionic and/or bosonic degrees of freedom. These are truly amazing results that confirm the generic strategy for constructing GOOFy-symmetric theories that we have developed here. The strategy could be formulated as follows. First we assume invariance with respect to generalized $C$ transformations for fields, than we allow independent transformations for Hermitian conjugated fields, which is then fixed by the requirement of invariance of kinetic terms. It should be remembered that coordinates  are rescaled as well ($x^\mu\to ix^\mu$) so that the independent transformation for Hermitian conjugated fields has to compensate that in order to make the kinetic terms invariant under GOOFy transformations.
 
{There is a comment here in order. 
We have just illustrated GOOFy transformations for many simple models varying numbers of fermionic and/or scalar fields. Note that dim-4 quartic scalar interactions were ignored there. The models were not viable candidates for Beyond the Standard Model (BSM) physics. They were chosen merely to show that invariance with respect to GOOFy transformations imply RGE stability of relations between entries of Yukawa matrices \eqref{con_gen} and mass parameters \eqref{Msquaregeneral} and/or \eqref{musquare}. 
It is worth emphasizing that, as mentioned above, the Hermitian conjugated fields transform non-consistently, i.e. for scalars as in \eqref{eq:trafo_rules} while for fermions as in \eqref{fer_trans}. It is tempting to argue that the GOOFy transformation is just a version of charge conjugation, however because of the non-consistent transformation of Hermitian conjugated fields GOOFy is different from $C$. 
Note also that the initial Yukawa couplings $\Gamma$ defined by \eqref{Yukawa_coupl} were unconstrained complex matrices, that became subject of restrictions \eqref{con_gen} when GOOFy invariance is required. Then we have shown that those restricted Yukawa matrices were RGE stable. 
In principle it would be possible to include dim-4 terms in the illustrative models discussed in sec.~\ref{Yukawa}. However, in order to limit unnecessary complications we have decided to include dim-4 only for the 2HDM, since it is a viable candidate for BSM physics. 
Symmetries of its scalar potential have been classified~\cite{Ivanov:2005hg,Ivanov:2006yq,Pilaftsis:2011ed}, it has been shown that there are six of them and relations between parameters 
implied by them are known.
In sec.~\ref{2HDM} we will consider the 2HDM and there we will show that invariance with respect to GOOFy transformations implies existence of RGE stable relations between parameters that are indeed new, they are not among those known from \cite{Ivanov:2005hg,Ivanov:2006yq,Pilaftsis:2011ed}. 
}
\section{GOOFy-invariant Standard Model}
\label{SM}
As it has been noticed earlier, for a single complex field $\Phi$ the $\Phi^\dagger M^2\Phi$ mass term is disallowed by a GOOFy symmetry. Since the SM  contains a single $SU(2)$ Higgs doublet, the model can not possess the Higgs mass parameter, making the model a non-viable candidate for a GOOFy theory of electroweak interactions. Nevertheless, we are going to discuss briefly constraints imposed upon Yukawa matrices to see if any other obstacles emerge. 
The Yukawa couplings of the SM reads
\bea
-\lcal_Y&=&\bar{q}_L\Gamma\Phi d_R + \bar{q}_L\Delta\tilde{\Phi} u_R + 
\bar{d}_R\Gamma^\dagger\Phi^\dagger q_L + \bar{u}_R\Delta^\dagger (\tilde{\Phi})^\dagger q_L,
\label{Ystart}
\eea
where $\tilde\Phi\equiv i\sigma_2\Phi^\star$ with $\sigma_2$ being the second Pauli matrix.
Adopting the generic strategy (described and tested earlier in various contexts), the GOOFy transformation shall be:
\bea
\Phi &\to X_\phi \Phi^\star,         \;\;\lsp\lsp  \Phi^\dagger &\to \Phi^T \bar X_\phi^\dagger, \label{GOOFy_SM}\\
q_L &\to -X_\alpha \gamma^0 C q_L^\star, \lsp \bar{q}_L &\to -q_L^T C^{-1} \bar X_\alpha^\dagger,\label{GOOFy_f1}\\
d_R &\to -X_\beta \gamma^0 C d_R^\star,  \lsp \bar{d}_R &\to -d_R^T C^{-1} \bar X_\beta^\dagger, \label{GOOFy_f2}\\
u_R &\to -X_\gamma \gamma^0 C u_R^\star, \lsp \bar{u}_R &\to -u_R^T C^{-1} \bar X_\gamma^\dagger \label{GOOFy_f3}.
\eea
where for the case of a single $SU(2)$ Higgs doublet $X_\phi = e^{i\theta}$ and $\bar X_\phi = - e^{i\theta}$, while $X_{\alpha,\beta,\gamma}$ and  $\bar X_{\alpha,\beta,\gamma}$ are unitary matrices acting in the $N_\psi$-dimensional flavour space. We will require (\ref{Ystart}) to be invariant under (\ref{GOOFy_SM}-\ref{GOOFy_f3}).

For simplicity we omit here the $\Delta$ terms. Then 
the conditions that follows from the requirement of invariance are
\bea
X_\beta^T\Gamma^T \bar X_\alpha^\star X_\phi&=&\Gamma^\dagger \label{Yuk_inv_1X},\\
X_\alpha^T\Gamma^\star \bar X_\beta^\star (\bar X_\phi)^\star&=&\Gamma . \label{Yuk_inv_2X}
\eea

From invariance of the kinetic fermionic terms one gets for $x_\mu \to ix_\mu$
\beq 
\bar X_{\alpha,\beta,\gamma} = -i X_{\alpha,\beta,\gamma}, 
\label{fer_kin} 
\eeq
while from the invariance of the scalar kinetic term we obtain
\beq 
\bar X_\phi=-X_\phi. \label{sca_kin} 
\eeq
Therefore, for GOOFy-invariant kinetic terms we get the following conditions
\bea
i X_\beta^T(e^{i\theta/2}\Gamma)^T X_\alpha^\star &=& (e^{i\theta/2}\Gamma)^\dagger, \label{fer_con1}\\
-i X_\alpha^T(e^{i\theta/2}\Gamma)^\star X_\beta^\star &=&(e^{i\theta/2}\Gamma). \label{fer_con2}
\eea
It is easy to see that the above equations are Hermitian conjugations of each other.

For 3 fermionic families we adopt for the GOOFy transformations the following  generalized $CP$ transformations found in \cite{Ecker:1987qp},
\bea
X_\psi= 
  \left( {\begin{array}{ccc}
   \cos\theta_\psi & \sin\theta_\psi & 0\\
  -\sin\theta_\psi & \cos\theta_\psi & 0\\
        0          &       0         & 1\\
  \end{array} } \right),
\eea
for $\psi=\alpha, \beta$ and $\gamma$. Non-trivial solutions of \eqref{fer_con1} - \eqref{fer_con2} may appear if the
determinant of the linear operator acting upon $\Gamma_{ab}$ in \eqref{fer_con1} - \eqref{fer_con2} vanishes, i.e. :
\bea
-256 \sin ^2(\alpha ) \sin ^2(\beta ) \sin ^2(\alpha -\beta ) \sin ^2(\alpha +\beta )=0
\eea
The non-trivial solutions with $0\leq\alpha,\beta\leq\pi/2$ are\\
{\it Solution 1:}
\bea
X_\alpha=X_\beta=\left(
\begin{array}{ccc}
	1 & 0 & 0 \\
	0 & 1 & 0 \\
	0 & 0 & 1 \\
\end{array}
\right)
,\quad
\Gamma=(1-i)e^{-i\theta/2}
\left(
\begin{array}{ccc}
	a_{11} & a_{12} & a_{13} \\
	a_{21} & a_{22} & a_{23} \\
	a_{31} & a_{32} & a_{33} \\
\end{array}
\right)
\eea
{\it Solution 2:}
\bea
X_\alpha=\left(
\begin{array}{ccc}
	1 & 0 & 0 \\
	0 & 1 & 0 \\
	0 & 0 & 1 \\
\end{array}
\right)
,\quad
X_\beta=\left(
\begin{array}{ccc}
	c_\beta  & s_\beta  & 0 \\
	-s_\beta & c_\beta & 0 \\
	0 & 0 & 1 \\
\end{array}
\right)
,\quad
\Gamma=(1-i)e^{-i\theta/2}
\left(
\begin{array}{ccc}
	0 & 0 & a_{13} \\
	0 & 0 & a_{23} \\
	0 & 0 & a_{33} \\
\end{array}
\right)
\eea
{\it Solution 3:}
\bea
X_\alpha=\left(
\begin{array}{ccc}
	c_\alpha  & s_\alpha  & 0 \\
	-s_\alpha & c_\alpha & 0 \\
	0 & 0 & 1 \\
\end{array}
\right)
,\quad
X_\beta=\left(
\begin{array}{ccc}
	1 & 0 & 0 \\
    0 & 1 & 0 \\
    0 & 0 & 1 \\
\end{array}
\right)
,\quad
\Gamma=(1-i)e^{-i\theta/2}
\left(
\begin{array}{ccc}
	0 & 0 & 0 \\
	0 & 0 & 0 \\
	a_{31} & a_{32} & a_{33} \\
\end{array}
\right)
\eea
{\it Solution 4:}
\bea
X_\alpha=X_\beta=\left(
\begin{array}{ccc}
	0 & 1 & 0 \\
	-1 & 0 & 0 \\
	0 & 0 & 1 \\
\end{array}
\right)
,\quad
\Gamma=e^{-i\theta/2}
\left(
\begin{array}{ccc}
	a_{11}+i b_{11} & a_{12}+i b_{12} & 0 \\
	i(a_{12}-i b_{12}) & -i(a_{11}-i b_{11}) & 0 \\
	0 & 0 & (1-i)a_{33} \\
\end{array}
\right)
\eea
{\it Solution 5:}
\bea
X_\alpha=X_\beta=\left(
\begin{array}{ccc}
	c_\alpha  & s_\alpha  & 0 \\
	-s_\alpha & c_\alpha & 0 \\
	0 & 0 & 1 \\
\end{array}
\right)
,\quad
\Gamma=(1-i)e^{-i\theta/2}
\left(
\begin{array}{ccc}
	a_{11} & a_{12} & 0 \\
	-a_{12} & a_{11} & 0 \\
	0 & 0 & a_{33} \\
\end{array}
\right)
\eea
where $a_{ij}$ and $b_{ij}$ are real numbers. Disregarding up and lepton Yukawa matrices, using results from PyR@TE~3, we find RGE-stability for all the above solutions up to two-loop order. The solution that could be viable (non-zero and non-degenerate masses ) is Solution~1, which unfortunately suffers from effectively not having complex phases and therefore no $CP$ violation. Concluding, both the scalar potential and Yukawa couplings of the SM are inconsistent with the GOOFy symmetry.

\section{Fermionic sector of the GOOFy-invariant 2HDM}
\label{2HDM}
{In this section, we are going to present the GOOFy-invariant 2HDM which constitutes a realistic BSM electroweak model with two Higgs doublets. The most general scalar potential for the 2HDM reads
\bea
V &=& m_{11}^2\Phi_1^\dagger\Phi_1+m_{22}^2\Phi_2^\dagger\Phi_2
-[m_{12}^2\Phi_1^\dagger\Phi_2+{\rm h.c.}]+\half\lambda_1(\Phi_1^\dagger\Phi_1)^2
+\half\lambda_2(\Phi_2^\dagger\Phi_2)^2
+\lambda_3(\Phi_1^\dagger\Phi_1)(\Phi_2^\dagger\Phi_2)\nonumber\\[8pt]
&&\quad
+\lambda_4(\Phi_1^\dagger\Phi_2)(\Phi_2^\dagger\Phi_1)
+\left\{\half\lambda_5(\Phi_1^\dagger\Phi_2)^2
+\big[\lambda_6(\Phi_1^\dagger\Phi_1)
+\lambda_7(\Phi_2^\dagger\Phi_2)\big]
\Phi_1^\dagger\Phi_2+{\rm h.c.}\right\}\,,
\label{eq:pot}
\eea
where, other than $m^2_{12}$ and $\lambda_{5,6,7}$, all parameters are real.
While the generic Yukawa Lagrangian for this model is as follows}
\bea
-\lcal_Y&=&\bar{q}_L\Gamma_a\Phi_a d_R + \bar{q}_L\Delta_a\tilde{\Phi}_a u_R + 
\bar{d}_R\Gamma_a^\dagger\Phi_a^\dagger q_L + \bar{u}_R\Delta_a^\dagger (\tilde{\Phi}_a)^\dagger q_L,
\label{start}
\eea
with $a=1,2$. GOOFy transformations for the two scalar doublets should have the same form as in \eqref{GOOFy_s} with $N_\phi=2$.
For the transformation of fermions we adopt \eqref{GOOFy_f1}-\eqref{GOOFy_f3} with $N_\psi=3$ fermionic families.
For the $\bar X$-matrices we adopt, as for the SM, \eqref{fer_kin} and \eqref{sca_kin}, while for unbared $X$-matrices
\bea
X_\phi=
  \left( {\begin{array}{cc}
   0 & e^{i\theta_1} \\
   e^{i\theta_2} & 0 \\
  \end{array} } \right),
\lsp
X_\psi= 
  \left( {\begin{array}{ccc}
   \cos\theta_\psi & \sin\theta_\psi & 0\\
  -\sin\theta_\psi & \cos\theta_\psi & 0\\
        0          &       0         & 1\\
  \end{array} } \right),
\label{eq:Xform}
\eea
for $\psi=\alpha, \beta$ and $\gamma$.

Explicitly, the transformations of the doublets read
\begin{alignat}{2}
	\Phi_1&\to e^{i\theta_1}\Phi_2^\star, &\quad
	\Phi_1^\dagger&\to -e^{-i\theta_1}\Phi_2^T,\nonumber\\
	\Phi_2&\to e^{i\theta_2}\Phi_1^\star, &\quad
	\Phi_2^\dagger&\to -e^{-i\theta_2}\Phi_1^T.
\end{alignat}
Note that the transformation considered here with $X_\phi$ given by \eqref{eq:Xform} is more general than the transformations obtained in \cite{Ferreira:2023dke}, they coincide for $\theta_1=0$ and $\theta_2=\pi$.
{Refraining from choosing a specific basis and demanding} invariance of the quadratic part of the potential under this transformation implies
\bea
m_{11}^2+m_{22}^2=0,\quad m_{12}^2(1+e^{i\Delta\theta})=0,
\eea
where $\Delta \theta\equiv \theta_1-\theta_2$ while the invariance of the quartic part of the potential under this transformation implies\footnote{Note that the last constraint implies $\lambda_6=\lambda_7=0$ unless $\Delta\theta=n\pi$.}
\bea
\lambda_2=\lambda_1,\quad \lambda_5(1-e^{2i\Delta\theta})=0,\quad \lambda_7=e^{-i\Delta\theta}\lambda_6=e^{i\Delta\theta}\lambda_6.
\eea
Comparing to the relations between potential parameters of the "original" GOOFy transformation found in ref.~\cite{Ferreira:2023dke}, we see that they differ in the relationship between $\lambda_6$, and $\lambda_7$, which actually depends on $\Delta\theta$. Only four options survive that correspond to different choices of $\Delta\theta$ and different relations between potential parameters, i.e.:
\bea
&&\Delta\theta=0:\quad m_{11}^2+m_{22}^2=0,\quad m_{12}^2=0,\quad \lambda_2=\lambda_1,\quad 
\lambda_7=\lambda_6,\label{eq:nzero},\\
&&\Delta\theta=\pi:\quad m_{11}^2+m_{22}^2=0,\quad \lambda_2=\lambda_1,\quad 
\lambda_7=-\lambda_6\label{eq:none},\\
&&\Delta\theta=\frac{2\pi}{3}:\quad m_{11}^2+m_{22}^2=0,\quad m_{12}^2=0,\quad \lambda_2=\lambda_1,\quad \lambda_5=\lambda_6=\lambda_7=0,\label{eq:ntwopithird}\\
&&\Delta\theta=\frac{4\pi}{3}:\quad m_{11}^2+m_{22}^2=0,\quad m_{12}^2=0,\quad \lambda_2=\lambda_1,\quad \lambda_5=\lambda_6=\lambda_7=0.
\label{eq:nfourpithird}
\eea
For $\Gamma_a$ terms, the conditions that follows from the requirement of GOOFy-invariance 
are~\footnote{
To impose the invariance conditions upon the $\Delta_a$ Yukawa matrices we would have to decide how to transform $\tilde\Phi_a$ and $\tilde\Phi_a^\dagger$. There exist two options for each of these that differ by sign:
\beq
\tilde\Phi_a=
\left\{ 
   \begin{array}{c}
   i\sigma_2 \left( \Phi_a^\dagger\right)^T \to -i\sigma_2 (X_\phi^\star)_{ab}\Phi_b\\
   i\sigma_2 \left( \Phi_a\right)^\star \to +i\sigma_2 (X_\phi^\star)_{ab}\Phi_b
  \end{array}
\right.
\nonumber 
\eeq
\beq
\tilde\Phi_a^\dagger=
\left\{ 
\begin{array}{c}
	-i \left( \Phi_a\right)^T\sigma_2 \to -i(\Phi_b^\dagger\sigma_2) (X_\phi)_{ab}\\
	-i \left( \Phi_a^\dagger\right)^\star\sigma_2 \to +i(\Phi_b^\dagger\sigma_2) (X_\phi)_{ab}
\end{array}
\right.
\nonumber 
\eeq
Thus, there are a total of four sign combinations to analyze. Two of these sign combinations will be discarded because they will imply $\Delta_a=0$. The two remaining sign combinations will lead to  
the invariance conditions for the $\Delta$-matrices: $ \Delta_b = \pm i X_\alpha^T \Delta_a^\star X_\gamma^\star (X_\phi)_{ab}$.
Note that the invariance conditions for $\Delta$'s differ from those for $\Gamma$'s only by a possible overall sign and by reversed sign of $\theta_{1,2}$.}

\bea
X_\beta^T\Gamma_a^T \bar X_\alpha^\star (X_\phi)_{ab}&=&\Gamma_b^\dagger \label{Yuk_inv_1}\\
X_\alpha^T\Gamma_a^\star \bar X_\beta^\star (\bar X_\phi)_{ab}^\star&=&\Gamma_b  \label{Yuk_inv_2}
\eea
The invariance conditions for the kinetic terms, $\bar X_\phi=-X_\phi$ and  $\bar X_\psi=-i X_\psi$ yields
\bea
\Gamma_1&=&-i X_\alpha^T\Gamma_2^\star X_\beta^\star e^{-i\theta_2},\label{eq:gamma1}\\
\Gamma_2&=&-i X_\alpha^T\Gamma_1^\star X_\beta^\star e^{-i\theta_1}\label{eq:gamma2}.
\eea
For $\Delta_a$ terms we find
\bea
\Delta_1&=&\pm i X_\alpha^T\Delta_2^\star X_\gamma^\star e^{i\theta_2},\label{eq:delta1}\\
\Delta_2&=&\pm i X_\alpha^T\Delta_1^\star X_\gamma^\star e^{i\theta_1}\label{eq:delta2}.
\eea
Inserting the expression for $\Gamma_2$ into the expression for $\Gamma1$ and the expression for $\Delta_2$ into the equation for $\Delta_1$ yields equations for $\Gamma_1$ and $\Delta_1$,
\bea
\Gamma_1&=&e^{-i\Delta\theta} (X_\alpha X_\alpha^\star)^\dagger \Gamma_1 (X_\beta X_\beta^\star), \label{eq:Gamma1}\\
\Delta_1&=&e^{+i\Delta\theta} (X_\alpha X_\alpha^\star)^\dagger \Delta_1 (X_\gamma X_\gamma^\star), \label{eq:Delta1}
\eea
for which we seek non-trivial solutions\footnote{We note that the equation for $\Delta_1$ has the same form as the equation for $\Gamma_1$, with the substitutions $\beta\to\gamma$ and $\Delta\theta\to-\Delta\theta$. We will not list the solutions for the $\Delta$-matrices, but they can be inferred from the solutions for the $\Gamma$-matrices with these substitutions along with (\ref{eq:delta2}).}. 
Without loss of generality we shall assume $0\leq\Delta\theta<2\pi$. Allowing for $\Gamma_1$ to be an arbitrary complex $3\times 3$ matrix, 
\bea
\Gamma_1&=&
  \left( {\begin{array}{ccc}
		z_{11} & z_{12} & z_{13}\\
		z_{21} & z_{22} & z_{23}\\
		z_{31} & z_{32} & z_{33}\\
\end{array} } \right),
\eea
with all $z_{ij}$ being complex parameters, eq.~\eqref{eq:Gamma1} represents a linear system of 18 equations with 18 unknown parameters. In order to find non-trivial solutions of this system of equations, we calculate the determinant of the resulting coefficient matrix of the system of equations, and demand that the determinant should vanish. The coefficient matrix can be written in block diagonal form, consisting of one $8\times8$ block, two $4\times4$ blocks and one $2\times2$ block. The determinant $D$ is then the product of the determinants of the four blocks,
\bea
D=D_1D_2D_3D_4,
\eea
where
\bea
D_1&=&256\sin ^2\left(\alpha +\beta -\frac{\Delta\theta}{2}\right) \sin ^2\left(\alpha -\beta +\frac{\Delta\theta}{2}\right)\\
&&\times\sin ^2\left(\alpha -\beta -\frac{\Delta\theta}{2}\right) \sin ^2\left(\alpha +\beta +\frac{\Delta\theta}{2}\right),\\
D_2&=&16\sin ^2\left(\alpha-\frac{\Delta\theta}{2} \right) \sin ^2\left(\alpha +\frac{\Delta\theta}{2}\right),\\
D_3&=&16\sin ^2\left(\beta-\frac{\Delta\theta}{2}\right) \sin ^2\left(\beta +\frac{\Delta\theta}{2}\right),\\
D_4&=&4\sin ^2\left(\frac{\Delta\theta}{2}\right).
\eea
Here, $D_1$ is the determinant of the block related to the parameters $\{z_{11},z_{12},z_{21},z_{22}\}$. $D_2$ is the determinant of the block related to the parameters $\{z_{13},z_{23}\}$. $D_3$ is the determinant of the block related to the parameters $\{z_{31},z_{32}\}$. $D_4$ is the determinant of the block related to the parameter $\{z_{33}\}$. The vanishing of each of these sub-determinants allows for non-zero values of the parameters related to said sub-determinant. 
We find several non-trivial solutions for $0\leq\alpha,\beta\leq\pi/2$.
We start by listing those solutions that correspond to $\Delta\theta=0$ and also allow for all quark masses to be non-zero.\\
\noi {\it Solution A:} ($D_1=D_2=D_3=D_4=0$),\quad $\Delta\theta=\alpha=\beta=0$,
\bea
\Gamma_1=
\left(
\begin{array}{ccc}
	z_{11} & z_{12} & z_{13} \\
	z_{21} & z_{22} & z_{23} \\
	z_{31} & z_{32} & z_{33} \\
\end{array}
\right), \quad 
\Gamma_2=-ie^{-i\theta_1}
\left(
\begin{array}{ccc}
	z_{11}^\star & z_{12}^\star & z_{13}^\star \\
	z_{21}^\star & z_{22}^\star & z_{23}^\star \\
	z_{31}^\star & z_{32}^\star & z_{33}^\star \\
\end{array}
\right)
\eea
{\it Solution D-1:} ($D_2\neq 0,\, D_3\neq0,\, D_1=D_4=0$),\quad $\Delta\theta=0,\, \beta=\alpha=\frac{\pi}{2}$,
\bea
\Gamma_1=
\left(
\begin{array}{ccc}
	z_{11} & z_{12} & 0 \\
	z_{21} & z_{22} & 0 \\
	0 & 0 & z_{33} \\
\end{array}
\right), \quad 
\Gamma_2=-ie^{-i\theta_1}
\left(
\begin{array}{ccc}
	z_{22}^\star & -z_{21}^\star & 0 \\
	-z_{12}^\star & z_{11}^\star & 0 \\
	0 & 0 & z_{33}^\star \\
\end{array}
\right)
\eea
{\it Solution D-2:} ($D_2\neq 0,\, D_3\neq0,\, D_1=D_4=0$),\quad $\Delta\theta=0,\, \beta=\alpha,\, \alpha\in(0,\frac{\pi}{2})$,
\bea
\Gamma_1=
\left(
\begin{array}{ccc}
	z_{11} & z_{12} & 0 \\
	-z_{12} & z_{11} & 0 \\
	0 & 0 & z_{33} \\
\end{array}
\right), \quad 
\Gamma_2=-ie^{-i\theta_1}
\left(
\begin{array}{ccc}
	z_{11}^\star & z_{12}^\star & 0 \\
	-z_{12}^\star & z_{11}^\star & 0 \\
	0 & 0 & z_{33}^\star \\
\end{array}
\right)
\eea
Next, we are listing all possible solutions corresponding to $\Delta\theta=\pi$ since this corresponds to the original GOOFy/$r_0$ transformation. We see that these solutions always implies at least one massless quark.\\
{\it Solution E:} ($D_1\neq 0,\, D_4\neq0,\, D_2=D_3=0$),\quad $\Delta\theta=\pi,\, \beta=\alpha=\frac{\pi}{2}$
\bea
\Gamma_1=
\left(
\begin{array}{ccc}
	0 & 0 & z_{13} \\
	0 & 0 & z_{23} \\
	z_{31} & z_{32} & 0 \\
\end{array}
\right), \quad 
\Gamma_2=-ie^{-i\theta_1}
\left(
\begin{array}{ccc}
	0 & 0 & -z_{23}^\star \\
	0 & 0 & z_{13}^\star \\
	-z_{32}^\star & z_{31}^\star & 0 \\
\end{array}
\right)
\eea
{\it Solution F:} ($D_3\neq 0,\, D_4\neq0,\, D_1=D_2=0$),\quad $\Delta\theta=\pi,\, \alpha=\frac{\pi}{2}$, $\beta=0$,
\bea
\Gamma_1=
\left(
\begin{array}{ccc}
	z_{11} & z_{12} & z_{13} \\
	z_{21} & z_{22} & z_{23} \\
	0 & 0 & 0 \\
\end{array}
\right), \quad 
\Gamma_2=-ie^{-i\theta_1}
\left(
\begin{array}{ccc}
	-z_{21}^\star & -z_{22}^\star & -z_{23}^\star \\
	z_{11}^\star & z_{12}^\star & z_{13}^\star \\
	0 & 0 & 0 \\
\end{array}
\right)
\eea
{\it Solution G:} ($D_2\neq 0,\, D_4\neq0,\, D_1=D_3=0$),\quad $\Delta\theta=\pi,\, \alpha=0$, $\beta=\frac{\pi}{2}$,
\bea
\Gamma_1=
\left(
\begin{array}{ccc}
	z_{11} & z_{12} & 0 \\
	z_{21} & z_{22} & 0 \\
	z_{31} & z_{32} & 0 \\
\end{array}
\right), \quad 
\Gamma_2=-ie^{-i\theta_1}
\left(
\begin{array}{ccc}
	-z_{12}^\star & z_{11}^\star & 0 \\
	-z_{22}^\star & z_{21}^\star & 0 \\
	-z_{32}^\star & z_{31}^\star & 0 \\
\end{array}
\right)
\eea
{\it Solution I:} ($D_1\neq 0,\, D_2\neq 0,\, D_4\neq0,\, D_3=0$),\quad $\Delta\theta=\pi$, $\beta=\frac{\pi}{2}$, $\alpha\in(0,\frac{\pi}{2})$,
\bea
\Gamma_1=
\left(
\begin{array}{ccc}
	0 & 0 & 0 \\
	0 & 0 & 0 \\
	z_{31} & z_{32} & 0 \\
\end{array}
\right), \quad 
\Gamma_2=-ie^{-i\theta_1}
\left(
\begin{array}{ccc}
	0 & 0 & 0 \\
	0 & 0 & 0 \\
	-z_{32}^\star & z_{31}^\star & 0 \\
\end{array}
\right)
\eea
{\it Solution J:} ($D_1\neq 0,\, D_3\neq 0,\, D_4\neq0,\, D_2=0$),\quad $\Delta\theta=\pi$, $\alpha=\frac{\pi}{2}$, $\beta\in(0,\frac{\pi}{2})$,
\bea
\Gamma_1=
\left(
\begin{array}{ccc}
	0 & 0 & z_{13} \\
	0 & 0 & z_{23} \\
	0 & 0 & 0 \\
\end{array}
\right), \quad 
\Gamma_2=-ie^{-i\theta_1}
\left(
\begin{array}{ccc}
	0 & 0 & -z_{23}^\star \\
	0 & 0 & z_{13}^\star \\
	0 & 0 & 0 \\
\end{array}
\right)
\eea
{\it Solution K:} ($D_1\neq 0,\, D_2\neq 0,\, D_3\neq0,\, D_1=0$),\quad $\Delta\theta=\pi$, $\beta=\frac{\pi}{2}-\alpha$, $\alpha\in(0,\frac{\pi}{2})$,
\bea
\Gamma_1=
\left(
\begin{array}{ccc}
	z_{11} & z_{12} & 0 \\
	z_{12} & -z_{11} & 0 \\
	0 & 0 & 0 \\
\end{array}
\right), \quad 
\Gamma_2=-ie^{-i\theta_1}
\left(
\begin{array}{ccc}
	-z_{12}^\star & z_{11}^\star & 0 \\
	z_{11}^\star & z_{12}^\star & 0 \\
	0 & 0 & 0 \\
\end{array}
\right)
\eea
Finally, we list two more solutions, corresponding to $\Delta\theta=\frac{2\pi}{3}$ and $\Delta\theta=\frac{4\pi}{3}$, respectively. These two solutions also allow for all quark masses being non-zero.\\
{\it Solution L:} ($D_1=D_2=D_3=0,\,D_4\neq0$),\quad $\Delta\theta=\frac{2\pi}{3},\, \alpha=\beta=\frac{\pi}{3}$,
\bea
\Gamma_1=
\left(
\begin{array}{ccc}
	z_{11} & iz_{11} & z_{13} \\
	iz_{11} & -z_{11} & -iz_{13} \\
	z_{31} & -iz_{31} & 0 \\
\end{array}
\right), \quad
\Gamma_2=-\frac{ie^{-i\theta_1}(\sqrt{3}+i)}{2}
\left(
\begin{array}{ccc}
	iz_{11}^\star & z_{11}^\star & -iz_{13}^\star \\
	z_{11}^\star & -iz_{11}^\star & z_{13}^\star \\
	-iz_{31}^\star & z_{31}^\star & 0 \\
\end{array}
\right)
\eea
{\it Solution M:} ($D_1=D_2=D_3=0,\,D_4\neq0$),\quad $\Delta\theta=\frac{4\pi}{3},\, \alpha=\beta=\frac{\pi}{3}$
\bea
\Gamma_1=
\left(
\begin{array}{ccc}
	z_{11} & -iz_{11} & z_{13} \\
	-iz_{11} & -z_{11} & iz_{13} \\
	z_{31} & iz_{31} & 0 \\
\end{array}
\right), \quad
\Gamma_2=-\frac{ie^{-i\theta_1}(\sqrt{3}-i)}{2}
\left(
\begin{array}{ccc}
	-iz_{11}^\star & z_{11}^\star & iz_{13}^\star \\
	z_{11}^\star & iz_{11}^\star & z_{13}^\star \\
	iz_{31}^\star & z_{31}^\star & 0 \\
\end{array}
\right)
\eea
{Note that only solutions A, D-1, D-2, L and M allow for all quarks to have non-zero masses. The remaining solutions will have at least one massless quark.} 
We are interested in checking the RGE-stability of the solutions listed here. Not only for the isolated Yukawa sector, but for a model where also the scalar potential with quadratic and quartic terms are present. There were four distinct relations between potential parameters, listed in eqs.~(\ref{eq:nzero} - \ref{eq:nfourpithird}). 
Combining the tree level constraints {upon potential parameters} with the tree level relations between the Yukawa matrices, (\ref{eq:gamma1}) and (\ref{eq:gamma2}), {we end up with the following four combined tree-level relations for $\Delta \theta=0,\pi,2\pi/3$ and $4\pi/3$.}\footnote{Note that the assumption of $m_{12}=0$ is only included here if $\Delta\theta\neq \pi$.}
\\
{\it Relation set 1 ($\Delta\theta=0$):}
\bea
&&m_{11}^2+m_{22}^2=0,\, m_{12}^2=0,\, \lambda_1=\lambda_2,\, \lambda_7=\lambda_6,\nonumber\\
&&\Gamma_1=-i X_\alpha^T\Gamma_2^\star X_\beta^\star e^{-i\theta_2},\,
\Gamma_2=-i X_\alpha^T\Gamma_1^\star X_\beta^\star e^{-i\theta_1},\nonumber\\
&&\Delta_1=\pm i X_\alpha^T\Delta_2^\star X_\gamma^\star e^{i\theta_2},\,
\Delta_2=\pm i X_\alpha^T\Delta_1^\star X_\gamma^\star e^{i\theta_1}.
\eea
{\it Relation set 2 ($\Delta\theta=\pi$):}
\bea
&&m_{11}^2+m_{22}^2=0,\, \lambda_1=\lambda_2,\, \lambda_7=-\lambda_6,\nonumber\\
&&\Gamma_1=-i X_\alpha^T\Gamma_2^\star X_\beta^\star e^{-i\theta_2},\,
\Gamma_2=-i X_\alpha^T\Gamma_1^\star X_\beta^\star e^{-i\theta_1},\nonumber\\
&&\Delta_1=\pm i X_\alpha^T\Delta_2^\star X_\gamma^\star e^{i\theta_2},\,
\Delta_2=\pm i X_\alpha^T\Delta_1^\star X_\gamma^\star e^{i\theta_1}.
\eea
{\it Relation set 3 ($\Delta\theta=\frac{2\pi}{3}$ or $\Delta\theta=\frac{4\pi}{3}$):}
\bea
&&m_{11}^2+m_{22}^2=0,\, m_{12}^2=0,\, \lambda_1=\lambda_2,\, \lambda_5=\lambda_6=\lambda_7=0,\nonumber\\
&&\Gamma_1=-i X_\alpha^T\Gamma_2^\star X_\beta^\star e^{-i\theta_2},\,
\Gamma_2=-i X_\alpha^T\Gamma_1^\star X_\beta^\star e^{-i\theta_1},\nonumber\\
&&\Delta_1=\pm i X_\alpha^T\Delta_2^\star X_\gamma^\star e^{i\theta_2},\,
\Delta_2=\pm i X_\alpha^T\Delta_1^\star X_\gamma^\star e^{i\theta_1}.
\eea
{Then solutions, A-M, obtained for Yukawas could be adopted above so that definite models are specified and could be investigated further. Using results from PyR@TE 3, we find RGE-stability up to two-loop order for {\it Relation set 1} for Solutions A, D-1 and D-2. For {\it Relation set 2}, we find RGE-stability up to two-loop order for Solutions E, F, G, I, J and K (which always imply at least one massless quark). For {\it Relation set 3}, we find RGE-stability up to two-loop order for Solutions L and M. Concluding we see that all solutions, A-M, have been confirmed RGE stable up to two-loop order. It is also worth noticing that the relations between potential parameters for {\it Relation Set 3} are identical to 0U(1) model listed in earlier work~\cite{Ferreira:2023dke}.}

{We note that the tree-level relations between potential parameters in Relation set 1, $m_{11}^2+m_{22}^2=0,\, m_{12}^2=0,\, \lambda_1=\lambda_2,\, \lambda_7=\lambda_6$ (with $\Delta \theta=0$) differs from the "original" relations for GOOFy/$r_0$ symmetry (with $\Delta \theta=\pi$) of the scalar potential, $m_{11}^2+m_{22}^2=0, \lambda_1=\lambda_2, \lambda_7=-\lambda_6$ that is part of Relation set 2 {(that implies at least one massless quark)}. Adopting covariance arguments under {\it basis transformation}, it has been shown in \cite{Ferreira:2023dke} that the relations $m_{11}^2+m_{22}^2=0, \lambda_1=\lambda_2, \lambda_7=-\lambda_6$ (with $\Delta\theta=0$) are stable under RGE to all orders when Yukawa couplings are neglected.
Here we observe analogous behaviour, i.e. RGE stability to all orders, for the relations  $m_{11}^2+m_{22}^2=0,\, m_{12}^2=0,\, \lambda_1=\lambda_2,\, \lambda_7=\lambda_6$ (with $\Delta \theta=0$). The reader is encouraged to follow details of the derivation in the Appendix.}

Since in the case of 2HDM, $z_{ab}$ are generic complex numbers there seems to be no issue concerning CP violation.
However there is an additional constraint that emerges from the requirement of absence of flavour changing neutral currents (FCNC). In other words the commutator $[\Gamma_1,\Gamma_2$] should be sufficiently small. This and other phenomenological issues will be discussed elsewhere.

\section{Conclusions}
\label{conc}
In this work the idea of GOOFy symmetries has been systematized and extended to fermionic degrees of freedom.
The main result of this paper is the formulation of the strategy for constructing GOOFy transformations inspired by generic $C$ transformation, followed by independent transformation of Hermitian conjugated fields and invariance of kinetic terms together with imaginary rescaling of coordinates.  All consequences of GOOFy-invariance that we are aware of are stable under the running of the RGE up to at least 2-loop order, in special cases up to 3-loop order.

Another important finding of this work is the generalization of GOOFy symmetries for fermions that enabled us to discuss constraints for Yukawa couplings, which again have been verified to be RGE stable up to 2 loops in all cases that were considered.

It has also been shown that even though the SM can not be made GOOFy invariant in a phenomenologically accepted manner (in particular with a non-zero Higgs vev), the 2HDM with its rich Yukawa sector allows for GOOFy constraints upon Yukawa couplings that are physically meaningful. 

Finally, a set of hitherto unknown relations between 2HDM potential parameters has been found,
\bea
m_{11}^2+m_{22}^2=0,\, m_{12}^2=0,\, \lambda_1=\lambda_2,\, \lambda_6=\lambda_7,
\eea
that has been shown to constitute a fixed point under the running of RGE up to at least three-loop order. {A preliminary study of implications of these relations show that they imply a 2HDM possessing a CP symmetry softly broken in a specific manner, i.e. by the condition $\mathrm{Tr} M^2=0$, that makes the model different from any other 2HDMs studied before.}

\section*{Acknowledgments}

{We would like to thank the referee for providing an elegant technical, alternative explanation for RGE invariant relations between parameters that we have noticed. That strategy is more conventional and, for instance, does not require imaginary scaling of coordinates. It is based upon spurious transformations of model parameters followed by the requirement of invariance of corresponding beta functions.
On the other hand, our GOOFy approach seems to be more systematic and therefore easier to apply when a comprehensive survey of possible RGE stable relations is needed. We would like to emphasize that it was a pleasure to interact with and learn from the referee.}

\noindent The work of B.G. is supported in part by the National Science Centre (Poland) as a research project 2023/49/B/ST2/00856.

\appendix
\section{Showing RGE-stability of relation set 1 to all orders}

{
The 2HDM potential can also  be written using the following bilinears \cite{Ferreira:2023dke},
\beq
\begin{array}{rcl}
	r_0 &=&
	\frac{1}{2}
	\left( \Phi_1^\dagger \Phi_1 + \Phi_2^\dagger \Phi_2 \right),
	\\*[2mm]
	r_1 &=&
	\frac{1}{2}
	\left( \Phi_1^\dagger \Phi_2 + \Phi_2^\dagger \Phi_1 \right)
	= \mbox{Re}\left( \Phi_1^\dagger \Phi_2 \right),
	\\*[2mm]
	r_2 &=&
	- \frac{i}{2}
	\left( \Phi_1^\dagger \Phi_2 - \Phi_2^\dagger \Phi_1 \right)
	= \mbox{Im} \left( \Phi_1^\dagger \Phi_2 \right),
	\\*[2mm]
	r_3 &=&
	\frac{1}{2}
	\left( \Phi_1^\dagger \Phi_1 - \Phi_2^\dagger \Phi_2 \right).
\end{array}
\label{eq:rs}
\eeq
as
\beq
V \,=\,M_\mu\,r^\mu\,+\,\Lambda_{\mu\nu}\,r^\mu\,r^\nu\,,
\eeq
where one uses Minkowski-like 4-vectors
\bea
r^\mu &=& (r_0\,,\,r_1\,,\,r_2\,,\,r_3) \,=\, (r_0\,,\,\vec{r})\,, \nonumber \\
M^\mu &=& \left(m^2_{11} + m^2_{22}\,,\, 2\mbox{Re}(m^2_{12})
\,,\, -2\mbox{Im}(m^2_{12})\,,\,m^2_{22} - m^2_{11}\right)\,=\, (M_0\,,\,\vec{M})\,,
\label{eq:defbil}
\eea
along with the tensor
\begin{align}
	\Lambda^{\mu\nu} & = \begin{pmatrix} \Lambda_{00} & \vec{\Lambda} \\
		\vec{\Lambda}^T & \Lambda \end{pmatrix}
	=
	\begin{pmatrix}
		\frac{1}{2}(\lambda_1 + \lambda_2) + \lambda_3 &
		-\mbox{Re}\left(\lambda_6 + \lambda_7\right) &
		\mbox{Im}\left(\lambda_6 + \lambda_7\right) &
		\frac{1}{2}(\lambda_2 - \lambda_1) \\
		-\mbox{Re}\left(\lambda_6 + \lambda_7\right) &
		\lambda_4 + \mbox{Re} \left( \lambda_5\right) &
		- \mbox{Im} \left( \lambda_5 \right) &
		\mbox{Re}\left(\lambda_6 - \lambda_7\right)
		\\
		\mbox{Im} \left( \lambda_6 + \lambda_7\right)&
		- \mbox{Im} \left( \lambda_5 \right) &
		\lambda_4 - \mbox{Re} \left( \lambda_5\right) &
		- \mbox{Im} \left( \lambda_6 - \lambda_7\right) \\
		\frac{1}{2}(\lambda_2 - \lambda_1) &
		\mbox{Re}\left( \lambda_6 - \lambda_7\right) &
		- \mbox{Im} \left( \lambda_6 - \lambda_7 \right) &
		\frac{1}{2}(\lambda_1 + \lambda_2) - \lambda_3
	\end{pmatrix}\,.
	\label{eq:Lambda}
\end{align}
It can be easily verified that the GOOFy transformation \eqref{eq:trafo_rules} with $X_\phi$ of the form \eqref{exphigeneral} for $\theta_1=\theta_2=0$ could be represented in the following simple manner in terms of the bilinears:
\bea
r_0\to-r_0,\quad r_1\to-r_1,\quad r_2\to -r_2,\quad r_3\to r_3\label{r0trafo}.
\eea
Decomposing the doublets as 
\bea
\Phi_1=
\begin{pmatrix}
	\phi_1+i\phi_2 \\
	\phi_3+i\phi_4
\end{pmatrix}, \quad
\Phi_2=
\begin{pmatrix}
	\phi_5+i\phi_6 \\
	\phi_7+i\phi_8
\end{pmatrix},
\eea
we find that the transformation 
\bea
\begin{pmatrix}
\phi_1\\
\phi_2\\
\phi_3\\
\phi_4\\
\phi_5\\
\phi_6\\
\phi_7\\
\phi_8
\end{pmatrix}
\to
\begin{pmatrix}
0 & 0 & 0 & 0 & 0 & -i & 0 & 0 \\
0 & 0 & 0 & 0 & -i & 0 & 0 & 0 \\
0 & 0 & 0 & 0 & 0 & 0 & 0 & -i \\
0 & 0 & 0 & 0 & 0 & 0 & -i & 0 \\
0 & -i & 0 & 0 & 0 & 0 & 0 & 0 \\
-i & 0 & 0 & 0 & 0 & 0 & 0 & 0 \\
0 & 0 & 0 & -i & 0 & 0 & 0 & 0 \\
0 & 0 & -i & 0 & 0 & 0 & 0 & 0
\end{pmatrix}
\begin{pmatrix}
\phi_1\\
\phi_2\\
\phi_3\\
\phi_4\\
\phi_5\\
\phi_6\\
\phi_7\\
\phi_8
\end{pmatrix}\label{transform}
\eea
is equivalent to (\ref{r0trafo}). }

{As it could have been anticipated the invariance of the potential $V$ with respect to the above transformation
implies relations between potential parameters which are contained in the {\it Relation set 1 $(\Delta\theta=0$)}:
\bea
m_{11}^2+m_{22}^2=0,\, m_{12}^2=0,\,\lambda_2=\lambda_1,\, \lambda_7=\lambda_6,
\eea
that, as we have shown earlier, are RGE-stable up to two loops even if Yukawa couplings are included.    Combined with 
\begin{gather}
B_\mu \to i B_\mu, \quad
W_{1\mu} \to i W_{1\mu},\quad
W_{2\mu} \to -i W_{2\mu},\quad
W_{3\mu} \to i W_{3\mu}.
\label{eq:igau}
\end{gather}
we also find invariance of the gauge- and kinetic terms of the Lagrangian.
}
{In order to argue that the combined constraints
\bea
m_{11}^2+m_{22}^2=0,\, m_{12}^2=0,\,\lambda_2=\lambda_1,\, \lambda_7=\lambda_6,
\eea
constitute a fixed point under the running of the RGE to all loop orders, we need to show that the form of $M_0$, $\vec{M}$, $\vec{\Lambda}$ and $\Lambda$ is preserved to all loop orders. The form of these quantities at the tree level are
\bea
M_0&=&0,\\
\vec{M}&=&(0,0,\times),\\
\vec{\Lambda}&=&(\times,\times,0),\\
\Lambda&=&
\begin{pmatrix}
	\times & \times & 0\\
	\times & \times & 0\\
	0 & 0 & \times
\end{pmatrix}.
\eea 
Using the results from Bednyakov \cite{Bednyakov:2018cmx}, in eq. (3.8) we find that $I_{0,1}=I_{1,3}=I_{2,3}=I_{3,3}=0$, which leads to $\beta_{M_0}=\beta_{m_{11}^2+m_{22}^2}=0$ to all orders.}

{The quantities listed in eq. (3.9), $\vec{M}, \Lambda\cdot \vec{M}, \Lambda^2\cdot\vec{M}$ all has the form $(0,0,\times)$, and since the quantity $I_M$ of eq. (3.10) vanishes, this implies $\beta_{\vec{M}}=(0,0,\times)$ to all orders, which again leads to $\beta_{m_{12}^2}=0$ to all orders.}

{Turning to eq. (3.5) and noting that $\vec{\Lambda}, \Lambda\cdot\vec{\Lambda}, \Lambda^2\cdot\vec{\Lambda}$ all are of the form $(\times,\times,0)$, we conclude that $\beta_{\vec{\Lambda}}=(\times,\times,0)$, which implies $\beta_{\lambda_2-\lambda_1}=0$ to all orders.}

{Finally, we turn to eq. (3.6), and notice that every term in this expansion is of the form 
\bea
\begin{pmatrix}
	\times & \times & 0\\
	\times & \times & 0\\
	0 & 0 & \times
\end{pmatrix},
\eea
which implies $\beta_{\lambda_7-\lambda_6}=0$ to all orders.}

{To summarize, we have shown that the combined constraints 
\bea
m_{11}^2+m_{22}^2=0,\, m_{12}^2=0,\,\lambda_2=\lambda_1,\, \lambda_7=\lambda_6,
\eea
constitute a fixed point under the running of the RGE to all loop orders.
}


\bibliography{biblio}
\bibliographystyle{JHEP}

\end{document}